\documentclass[]{emulateapj}

\usepackage{amsmath}
\usepackage{dcolumn,multirow,ccaption}

\usepackage{graphicx}
\usepackage{color}
\usepackage{rotating}
\usepackage{ccaption}
\usepackage{lscape}

\def\smppc2{${\rm M}_{\odot} {\rm pc}^{-2}$}		
\def\ncm3{cm$^{-3}$}						
\def\kms{$\rm km\;s^{-1}$}		

\shorttitle{Large Peculiar Motions of Star-Forming Regions}
\shortauthors{Baba et al.}

\begin{document}

\title{The origin of large peculiar motions of star-forming regions and
spiral structures of our Galaxy}

\author{Junichi \textsc{Baba}\altaffilmark{1}, 
        Yoshiharu \textsc{Asaki}\altaffilmark{2,3}, 
        Junichiro \textsc{Makino}\altaffilmark{1,4,5}, 
        Makoto \textsc{Miyoshi}\altaffilmark{6},
        Takayuki \textsc{R.Saitoh}\altaffilmark{1,7},
        Keiichi \textsc{Wada}\altaffilmark{1,4,5}
}

\altaffiltext{1}{Division of Theoretical Astronomy, National Astronomical
Observatory of Japan, 2--21--1 Osawa, Mitaka-shi, Tokyo 181--8588.}

\altaffiltext{2}{Institute of Space and Astronautical Science, 3--1--1
Yoshinodai, Sagamihara, Kanagawa 229--8510, Japan.}

\altaffiltext{3}{Department of Space and Astronautical Science, School of
Physical Sciences, The Graduate University for Advanced Studies (SOKENDAI),
3--1--1 Yoshinodai, Sagamihara, Kanagawa 229--8510, Japan.}

\altaffiltext{4}{Department of Astronomical Science, School of Physical
Sciences, The Graduate University for Advanced Studies (SOKENDAI), 2--21--1
Osawa, Mitaka-shi, Tokyo 181--8588, Japan.}

\altaffiltext{5}{Center for Computational Astrophysics, National
Astronomical Observatory of Japan, 2--21--1 Osawa, Mitaka, Tokyo 181-8588,
Japan.}

\altaffiltext{6}{Division of Radio Astronomy, National Astronomical
Observatory of Japan, 2--21--1 Osawa, Mitaka, Tokyo 181-8588, Japan.}

\altaffiltext{7}{Japan Society for the Promotion of Science for Young
Scientists Research Fellowship for Young Scientists.}

\email{baba.junichi@nao.ac.jp}

\begin{abstract}
Recent Very Long Baseline Interferometer (VLBI) observations determined the distances and proper motions of star-forming regions in spiral arms directly. They showed that star-forming regions and young stars have large peculiar motions, as large as 30 ${\rm km}~{\rm s}^{-1}$ with complex structures. Such a large peculiar motion is incompatible with the prediction of the standard theory of quasi-stationary spiral arms. We use a high-resolution, self-consistent $N$-body+hydrodynamical simulation to explore how the spiral arms are formed and maintained, and how star-forming regions move. We found that arms are not quasi-stationary but transient and recurrent, as suggested in alternative theories of spiral structures. Because of this transient nature of the spiral arms, star-forming regions exhibit a trend of large and complex non-circular motions, which is qualitatively consistent with the VLBI observations. Owing to this large non-circular motion, a kinematically estimated gas map of our Galaxy has a large systematic errors of $\sim 2-3$ kpc in the distance from the Sun.
\end{abstract}


\section{Introduction}

How the spiral arms in a disk galaxy are maintained has been a long-standing mystery. Standard understanding is that stellar spiral arms are maintained as stationary density waves \citep{LinShu1964,BertinLin1996}.  In this theory, the interstellar medium (ISM), whose motion is slightly perturbed by these spiral arms, forms visible spiral arms of star-forming regions and young stars.  Alternative theories, in which stellar spiral arms are transient and recurrent \citep[e.g.][]{GoldreichLynden-Bell1965, JulianToomre1966, Toomre1981,SellwoodCarlberg1984}, have been also proposed, and there has been no conclusive observational evidence or theoretical argument on which theory is right.  

\begin{figure*}[htbp]
\begin{center} 
	\includegraphics[width=.4\textwidth]{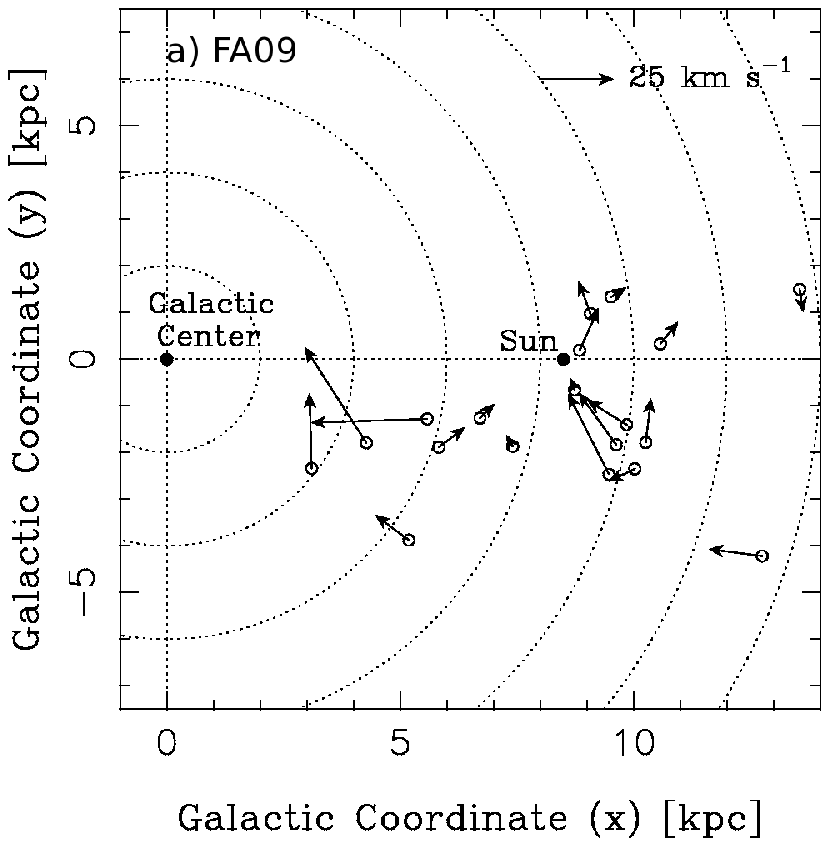}	
	\includegraphics[width=.4\textwidth]{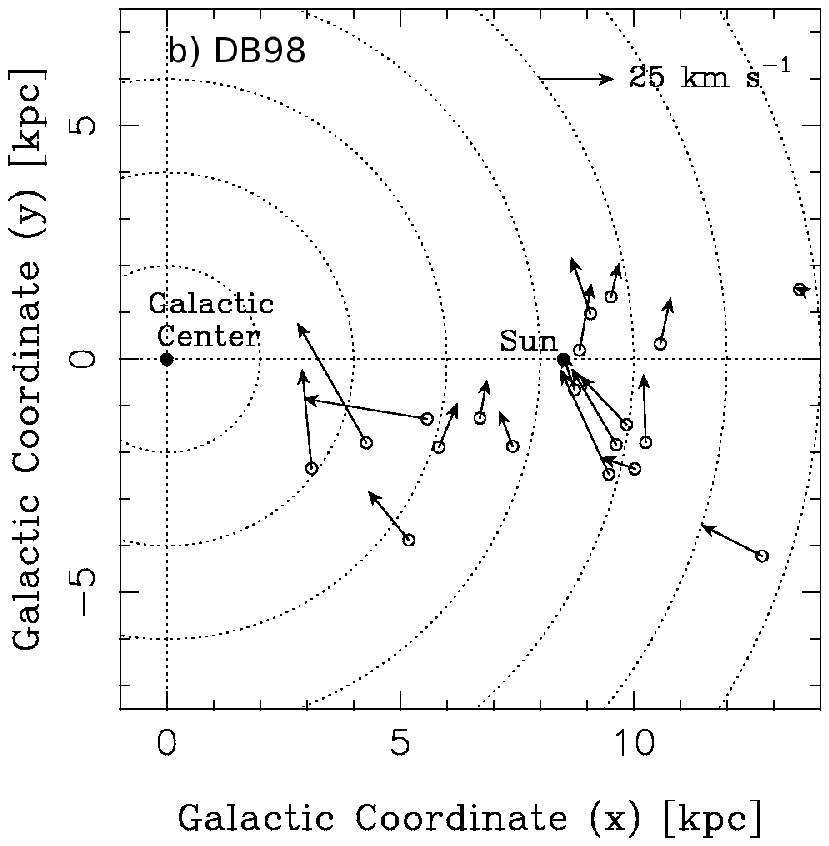}	
	\caption{Spacial distribution and peculiar velocities of the star-forming regions in our Galaxy (see Table 1). a) Peculiar velocities of star-forming regions in our Galaxy, based on the solar motion relative to the LSR of FA09. b) Same as panel a), but the solar motion relative to the LSR of DB98.}
	\label{fig:VLBIobs}
\end{center}    	
\end{figure*}

Recent Very Long Baseline Interferometer (VLBI) observations \citep[][and references therein]{Reid+2009b} determined the distances and proper motions of star-forming regions in spiral arms of our Galaxy directly. The distance and the three-dimensional motion of a star-forming region W3 OH in the Perseus spiral arm have been determined through the parallax measurement using the Very Long Baseline Array (VLBA) \citep{Xu+2006}. Its distance from the Sun is $1.95 \pm 0.05~{\rm kpc}$ and its peculiar motion (i.e. deviations from the circular rotation) is $ (U, V, W) = (17 \pm 1, -14 \pm 1, -0.8 \pm 0.5)~{\rm km}~{\rm s}^{-1}$, where the three components are the residual from the circular velocity in the direction of the galactic center ($U$), the direction of rotation ($V$), and the direction perpendicular to the galactic plane ($W$) 
\footnote{In \citet{Xu+2006}, the solar motion relative to the local standard of rest (LSR) is assumed to be $(U_\odot, V_\odot, W_\odot) = (10.0, 5.25, 7.17)~{\rm km}~{\rm s}^{-1}$ \citep{DehnenBinney1998}, and the galactic radius and circular rotation of the LSR are assumed to be $R_0 = 8.5~{\rm kpc}$ and $\Theta_0 = 220~{\rm km}~{\rm s}^{-1}$.}. The two components in the galactic plane sum up to $22~{\rm km}~{\rm s}^{-1}$, which is about 10\% of the circular velocity of the solar system, $\Theta_0 = 220~{\rm km~s^{-1}}$ \citep{KerrLyndelBell1986}.  W3 OH is not an exceptional object. Table \ref{tbl:VLBIobs} and Figure \ref{fig:VLBIobs} show the compilation of the results of recent high-accuracy astrometric observations of distances and peculiar velocities of star forming regions using the VLBA and VLBI Exploration of Radio Astrometry (VERA) \citep{Xu+2006, Asaki+2007, Honma+2007, Choi+2008, Kim+2008, Sato+2008, Zhang+2009, Moellenbrock+2009, Hachisuka+2009, Reid+2009a, Moscadelli+2009, Xu+2009, Brunthaler+2009, Bartkiewicz+2008}.  Most sources have peculiar velocities larger than $10~{\rm km}~{\rm s}^{-1}$ with various directions.

These large peculiar motions of star-forming regions are incompatible with the standard understanding of the spiral arm based on the stationary density wave theory \citep{LinShu1964} and a standing galactic shock solution in a tight-winding spiral potential \citep{Fujimoto1968, Roberts1969, RobertsYuan1970}, which predicts non-circular flow depending on the strength of spiral potential, whose amplitude is $\sim 8-12~{\rm km~s^{-1}}$ even for a very strong spiral \citep[i.e. $F=5$\%][]{ShuRoberts1973}. More importantly, the ``galactic shock'' solutions predict that the spiral perturbation can only generate a laminar flow (i.e. streamline flow). This picture does not match the observations.

It has been recognized that ISM in external barred spiral galaxies, such as M51, show large non-circular motions \citep[e.g.][]{Garcia-Burillo+1993}. This has been also confirmed by hydrodynamic simulations of an isothermal gas in a bar potential \citep[e.g.][]{Athanasoula1992}. Peculiar features of the longitude-velocity diagrams of atomic and molecular ISM in our Galaxy are also interpreted as non-circular flows under the influence of the stellar bar \citep{Binney+1991, Combes1991, Rodriguez-FernandezCombes2008}.  However, almost all previous numerical studies of gas dynamics in a barred-spiral potential were not realistic enough to be directly compared with observations by the VLBI, because star-forming dense gas clouds were not self-consistently treated in their models.  Either an isothermal equation of state with gas temperature of $10^4$ K was assumed \citep{FriedliBenz1993}, or the ISM was represented by collisional particles with a mass spectrum \citep{Tomisaka1986, Garcia-Burillo+1993}. \citet{WadaKoda2001} clearly showed how the structures of the ISM differ in the isothermal and the multi-phase gas models in a bar potential.

Recently, simulations of gas disks in a fixed (i.e. time-independent) background stellar potential were tried with very high spatial resolutions, both with an Euler grid technique \citep{WadaNorman2007} and a smoothed particle hydrodynamics (SPH) technique \citep{Saitoh+2008}.  Thanks to the high spatial and mass resolutions, they were able to follow the cooling of gas to quite low temperature (below $100~{\rm K}$), and thus found that gas disks develop complex fractal-like structures. In these simulations, quasi-steady distribution of gas was maintained through the balance between the two competing processes. One is the structure formation through radiative cooling and gravitational contraction, and the other is the disruption through tidal interactions, heating by supernova (SN) explosion and shear due to a differential rotation. However, no clear spiral structure was formed. This is not surprising because the stellar disk is expressed by a fixed, axisymmetric potential field or by $N$-body particles without a sufficient resolution.

In the present paper, we combine high-resolution $N$-body and hydrodynamic calculations in order to understand the peculiar motion of star forming regions observed in our Galaxy.  In \S \ref{sec:method}, we describe our numerical model, in which a quasi-steady, isolated `spiral' galaxy in a fixed spherical potential is produced from a featureless stellar and gas disk. The evolution of the stellar disk, and distribution and kinematics of cold gas and young stars are presented in \S \ref{sec:results}. This section also compares peculiar velocities of star-forming gas in our simulation with those of observations. In \S \ref{sec:discussion}, we discuss the rotation speed of the star-forming gas, origins of large peculiar motions of star-forming gas, and how kinematic distances of the Galactic objects differs from the true ones. The results are summarised in \S \ref{sec:sum}.

\begin{table*}
	\caption{Peculiar motions (deviation from a galactic circular rotation) of star forming regions in our Galaxy measured by VLBI astrometric observations of ${\rm H}_2{\rm O}$ and ${\rm CH}_3{\rm OH}$ masers obtained by the VLBA and VERA. $D$ [kpc] is distance from the Sun, $U$, $V$, and $W$ [\kms] are three components of peculiar motions. $U$ is toward the Galactic center, $V$ is toward Galactic rotation, and $W$ is toward the North Galactic Pole. Galactic radius and circular rotation of the LSR are assumed to be $R_0 = 8.5~{\rm kpc}$ and $\Theta_0 = 220~{\rm km}~{\rm s}^{-1}$. The flat rotation curve is assumed. The solar motion relative to the LSR is $(U_\odot, V_\odot, W_\odot) = (7.5, 13.5, 6.8)~{\rm km}~{\rm s}^{-1}$ \citep[][hereafter FA09]{FrancisAnderson2008} and $(U_\odot, V_\odot, W_\odot) = (10.00, 5.25, 7.17)~{\rm km}~{\rm s}^{-1}$ \citep[][hereafter DB98]{DehnenBinney1998}.}
	\begin{center}
	\begin{tabular}{lrrrrrrrrl}
	\hline
										&	& \multicolumn{3}{c}{FA09} && \multicolumn{3}{c}{DB98} &  \\
	\cline{3-5} \cline{7-9}
	Source						 & D 		& \multicolumn{1}{r}{U}&\multicolumn{1}{r}{V} &\multicolumn{1}{r}{W} 
										&			& \multicolumn{1}{r}{U}&\multicolumn{1}{r}{V} &\multicolumn{1}{r}{W} & Reference \\
	\hline
	W3 OH           			& 1.95 		& 14.4 	&   -6.3 &   0.7 		&& 18.0 &  -14.2 &  1.1 & \cite{Xu+2006}\\
	IRAS 00420+5530	& 2.17 	& 15.8 	&  -15.1 &  -2.2 		&&  19.8 &  -22.7 &  -1.9 & \cite{Moellenbrock+2009} \\
	WB89-437        		& 6.0  	& 18.0 	&    3.5 &   0.5 		&&  23.0 &   -3.5 &   0.8 & \cite{Hachisuka+2009} \\
	S 252           			& 2.10 	& -6.3 	&   -7.3 &  -2.4 		&& -4.0 &  -15.6 &  -2.0 & \cite{Reid+2009a} \\
	G232.6+1.0				& 1.68 & -5.6 	&   -2.4 &   0.4 		&& -4.2 &  -11.0 &   0.8 & \cite{Reid+2009a}\\
	Cep A						& 0.70	& 1.9 	&   -4.0 &  -5.8 		&& 5.0 &  -12.0 &  -5.4 & \cite{Moscadelli+2009}  \\
	NGC 7538        		& 2.65 &  20.4 &  -23.0 & -11.3 	&& 24.9 &  -30.4 & -10.9 & \cite{Moscadelli+2009} \\
	G59.7+0.1       		& 2.16 &   3.0 &   -3.2 &  -4.6 		&& 7.5 &  -10.6 &  -4.3 & \cite{Xu+2009} \\
	W 51 IRS2       		& 5.10 	&  14.2 &    0.0 &  -3.4 		&& 21.2 &   -5.1 &  -3.0 & \cite{Xu+2009} \\
	G35.20-0.74     	& 2.19 	&  -3.9 &   -5.6 &  -8.9 		&& 0.1 &  -13.3 &  -8.5 & \cite{Zhang+2009}\\
	G35.20-1.74     	& 3.27 	&  -6.3 &   -9.0 &  -9.7 		&& -1.4 &  -16.1 &  -9.4 & \cite{Zhang+2009}\\
	G23.01-0.41     	& 4.59 	&  32.0 &  -22.0 &  -1.7 		&& 37.5 &  -28.6 &  -1.4 & \cite{Brunthaler+2009}\\
	G23.44-0.18     	& 5.88 	&  16.0 &  -20.0 &   1.5 	&& 23.0 &  -25.1 &   1.9 & \cite{Brunthaler+2009}\\
	G23.657-00.127  & 3.19 	&  38.4 &   10.2 &   3.8 	&& 42.7 &    2.7 &   4.1 & \cite{Bartkiewicz+2008}  \\
	S 269           			& 5.28 &  -0.3 &    7.7 &  -4.9 	&& 1.3 &   -0.8 &  -4.5 & \cite{Honma+2007} \\
	NGC281 West     	& 2.82 &   7.7 &    6.3 & -13.9 	&& 12.0 &   -1.2 & -13.5 & \cite{Sato+2008} \\
	S Per           			& 2.51 	&   0.8 &  -15.1 & -10.3 	&& 4.7 &  -22.8 & -10.0 & \cite{Asaki+2007}  \\
	VY CMa          		& 1.14 	&   2.9 &  -11.2 &  -3.8  		&& 4.5 &  -19.7 &  -3.5 & \cite{Choi+2008} \\
	Orion KL        		& 0.42 	&  -6.8 &  -14.3 &  10.5  	&& -4.5 &  -22.6 &  10.9 & \cite{Kim+2008} \\  
	\hline
	\end{tabular}
	\end{center}
	\label{tbl:VLBIobs}
\end{table*}

\section{Methods and Model Setup}
\label{sec:method}

\subsection{Numerical Methods}
In our model, evolution of a stellar disk with the multi-phase gas in a halo potential, taking into account star formation from cold, dense gas and energy feedback from SNe, was solved numerically.  We used our original $N$-body/gas simulation code {\tt ASURA} \citep{Saitoh+2008,Saitoh+2009}. Hydrodynamics was solved by the standard SPH methods and the smoothing length was allowed to vary both in space and time with the constraint that the typical number of neighbor each particle is near $N_{\rm nb} = 32$. The artificial viscosity term \citep{Monaghan1997} and the correction term to avoid large entropy generation in pure shear flows \citep{Balsara1995} were used. Self-gravity of stars and SPH particles were calculated by the Tree with GRAPE method \citep{Makino1991} which is a combination of the Tree method \citep{BarnesHut1986} and GRAPE \citep{Sugimoto+1990}. The opening angle was set to be 0.5. In these simulations, we adopted a software emulator of GRAPE, named Phantom-GRAPE (Nitadori et al. in prep.). Phantom-GRAPE is tuned by Single Instruction Multiple Data (SIMD) instructions which are equipped by modern CPUs. The leapfrog time-integrator was adopted.

Metallicity-dependent radiative cooling of the gas was solved assuming an optically thin cooling function with solar metallicity that covered a wide range of temperature, $20$ K through $10^8$K \citep{WadaNorman2001}. Uniform photo-electric heating of dust by the far-ultraviolet radiation (FUV) observed  in the solar neighborhood was taken into account \citep{Wolfire+1995, GerritsenIcke1997}. 

The treatment of star formation and the heating due to the SN feedback was the same as in \citet{Saitoh+2008}. We adopted the single stellar population approximation, with the Salpeter initial mass function \citep{Salpeter1955} and the mass range of $0.1-100 M_{\odot}$. An SPH particle was replaced with a star particle following the Schmidt law \citep{Schmidt1959} with a local star formation efficiency ($C_{\ast} = 0.033$) in a probabilistic manner, if criteria (1) $n_{\rm H} > 100$ cm$^{-3}$, (2) $T_{\rm g} < 100$ K, and (3) $\nabla\cdot\vec{v} < 0$ are satisfied. Here we assumed the dense and cold gas as the potential site of star formation. Advantages of this set of star formation criteria are as follows \citep{Saitoh+2008}: (1) simulations reproduce realistic structures of the ISM and young stars, while the set of criteria ($n_{\rm H} > 0.1$ cm$^{-3}$ , $T_{\rm g} < 15000$ K, and $\nabla \cdot \vec{v} < 0$), adopted in many previous studies of galaxy formation \citep[e.g.,][]{Governato+2008} do not reproduce those structures, and (2) global (galactic) star formation rate is not directly proportional to the local star formation efficiency,  $C_{\ast}$, but largely controlled by the global evolution of the ISM from reservoir to dense gas where stars are formed \citep[for details, see \S 5.2 in][]{Saitoh+2008}. The second point is specially important for numerical simulations of galaxy formation and evolution since our result does not depend strongly on the value of $C_{\ast}$ used. We implemented type-II SN feedback, where the energy from SNe is injected to the gas around the star particles in the thermal energy.  Each SN releases $10^{51}~{\rm ergs}$ of thermal energy for the surrounding ISM.

\subsection{Initial Model}

The procedure to prepare the initial model is as follows.  We first place a pure $N$-body stellar disk of an exponential profile in the center of a static dark-matter (DM) halo potential. Details of the halo potential and the disk model will be given in \S \ref{sec:model}. Then, we let the stellar disk evolve for $2~{\rm Gyr}$.  After a bar and stellar spirals have developed well, we added a gas component to the stellar disk (\S 2.2.2). We used this combination of fixed DM potential and stellar and gas disks as the initial condition of the simulation for the gas disk-stellar disk-static halo model.

\subsubsection{DM Halo + Stellar Disk Model \label{sec:model}}
We adopted a fixed, spherical DM halo as a host of an $N$-body stellar disk. The density profile of the DM halo follows the Navarro-Frenk-White (NFW)-profile:
\begin{equation}
\rho_{\rm h}(r)=\frac{\rho_0}{r/r_{\rm s}(1+r/r_{\rm s})^2},
\end{equation}
where
\begin{equation}
\rho_0 = \frac{M_{\rm h}}{4\pi {R_{\rm h}}^3}
\frac{{C_{\rm NFW}}^3}{\ln(1+C_{\rm NFW})+C_{\rm NFW}/(1+C_{\rm NFW})},
\end{equation}
\begin{equation}
  C_{\rm NFW} = R_{\rm h}/r_{\rm s},
\end{equation}
$C_{\rm NFW}$, $M_{\rm h}$, and $R_{\rm h}$ are the concentration parameter of the DM halo, the mass, and the virial radius, respectively.  The concentration parameter is set to be $C_{\rm NFW} = 5$. The mass is $M_{\rm h}=6.3 \times 10^{11}~M_{\odot}$ and the virial radius is $R_{\rm h} = 122~{\rm kpc}$.  The values of $M_{\rm h}$ and $R_{\rm h}$ are determined by using the spherical collapse model \citep[Equation (2) in][] {Mo+1998} with a circular velocity of the halo $V_{\rm c} = 150~{\rm km}~{\rm s}^{-1}$ and a formation redshift of the halo $z_{\rm f} = 1$.

The initial condition for the stellar disk is generated using Hernquist's method \citep{Hernquist1993}. The density profile of the stellar disk is given by 
\begin{equation}
\rho_{\rm sd}(R,z) = \frac{M_{\rm sd}}{4\pi {R_{\rm sd}}^2 z_{\rm sd}}
\exp(-R/R_{\rm sd}) {\rm sech}^2(z/z_{\rm sd}),\label{eq:stellardisk}
\end{equation}
where $M_{\rm sd}$ is the mass of the stellar disk, $R_{\rm sd}$ is the radial scale-length, and $z_{\rm sd}$ is the vertical scale-length, respectively. The vertical velocity dispersion $\sigma_{\rm z}$ is related to the surface stellar density $\Sigma_{\rm sd}$ and $z_{\rm sd}$:
\begin{equation}
	\sigma_{\rm z}^2(R) = \pi G z_{\rm sd}\Sigma_{\rm sd}(R),
\end{equation}
where $G$ is the gravitational constant. The radial velocity dispersion $\sigma_{\rm R}$ is also assumed to be directly related to the surface density:
\begin{equation}
	\sigma_{\rm R}^2(R) = A\exp[-R/R_{\rm sd}].
\end{equation}
The normalization constant $A$ is fixed in such a way that at some reference radius $R_{\rm ref} (= 1.5R_{\rm sd})$ the radial random velocities are $Q_{\rm ref}$ times the critical value needed to stabilize a differentially rotating disk against local perturbations:
\begin{equation}
	\sigma_{\rm R}(R_{\rm ref}) = Q_{\rm ref}\frac{3.36G\Sigma_{\rm sd}(R_{\rm ref})}{\kappa(R_{\rm ref})},
\end{equation}
where $\kappa(R_{\rm ref})$ is the epicyclic frequency at $R_{\rm ref}$. The azimuthal component of the rotational velocity ($V_{\phi}$) and its dispersion ($\sigma_{\phi}$) are found using the asymmetric drift and the epicyclic approximations:
\begin{eqnarray}
	&&V_{\phi}^2 =  V_{\rm cir}^2 - \sigma_{\rm R}^2\left(\frac{2R}{R_{\rm sd}} + \frac{\kappa^2}{4\Omega^2} - 1\right), \\
	&&\sigma_{\phi}^2 = \sigma_{\rm R}^2 \frac{\kappa^2}{4\Omega^2}.
\end{eqnarray}
where $V_{\rm cir}$ and $\Omega \equiv V_{\rm cir}/R$ are the circular velocity and the angular frequency at a given $R$, respectively. The model parameters of the DM halo and the $N$-body stellar disk are summarized in Table \ref{tbl:halodiskmodel}.

As suggested by many previous studies \citep[e.g.][]{OstrikerPeebles1973, Efstathiou+1982}, a massive disk experiences a bar instability.According to \citet{Efstathiou+1982}, who used $N$-body simulations to investigate bar instabilities of exponential disks embedded in fixed DM haloes, the bar instability takes place in the stellar disks when the criterion
\begin{equation}
	\epsilon_{\rm b} \equiv \frac{V_{\rm max}}{\sqrt{GM_{\rm sd}/R_{\rm sd}}} \lesssim 1.1 
\end{equation}
is satisfied. Here $V_{\rm max}$ is the maximum rotation velocity of the disk. In our model, the value of $\epsilon_{\rm b}$ is $\simeq 0.8$ for the initial state. Thus our model is unstable against bar formation.

\begin{table}
	\caption{Model parameters of DM halo + stellar disk}
	\begin{center}
	\begin{tabular}{ll}
	\hline
		Total mass of halo ($M_{\rm h}$)                &$6.3\times 10^{11} M_{\odot}$ \\
		Virial radius of halo ($R_{\rm h}$)                & $122~{\rm kpc}$ \\
		Concentration parameter ($C_{\rm NFW}$)& $5.0$ \\
	 	Total mass of stellar disk ($M_{\rm sd}$)	& $3.2\times 10^{10} M_{\odot}$ \\
		Scale length of stellar disk ($R_{\rm sd}$) & ${3.0}~{\rm kpc}$ \\
		Scale height of stellar disk ($z_{\rm sd}$)	& ${0.30}~{\rm kpc}$ \\
		Toomre's $Q$-value at $R=1.5~R_{\rm sd}$ ($Q_{\rm ref}$) & ${1.0}$ \\
	\hline
	\end{tabular}
	\end{center}
	\label{tbl:halodiskmodel}
\end{table}%

\subsubsection{DM Halo + Stellar Disk + Gas Disk model}
We added a gas component to the barred spiral stellar disk in which global quantities such as the radial surface density profile, $Q$ parameter, and velocity dispersions did not significantly change. The gas disk has a radially exponential density profile, whose total mass is set to $M_{\rm gd} = 0.1~M_{\rm sd}$, and radial and vertical scale-lengths of the gas disk are set to $R_{\rm gd} = 2~R_{\rm sd}$ and $z_{\rm gd} = 0.2~{\rm kpc}$, respectively.  The gas disk initially has the same circular velocity determined by the mass distribution. Although the velocity dispersion of $10~{\rm km~s^{-1}}$ is added for the vertical direction, there is no effect on our results. The initial temperature of gas is set to be $10^4~{\rm K}$. Model parameters of the gas disk are summarized in Table \ref{tbl:gasmodel}.

\begin{table}[htdp]
	\caption{Model parameters of the gas disk}
	\begin{center}
	\begin{tabular}{ ll }
	\hline
 	Total mass ($M_{\rm gd}$)				&$3.2\times 10^{9} M_{\odot}$ \\
	Scale length ($R_{\rm gd}$)				&${6.0}~{\rm kpc}$ \\
	Scale height ($z_{\rm gd}$)				&$0.2~{\rm kpc}$ \\
	Initial vertical velocity dispersion	&$10~{\rm km~s^{-1}}$\\
	Initial temperature ($T_{\rm in}$)	&$10^4~{\rm K}$ \\
	\hline
	\end{tabular}
	\end{center}
	\label{tbl:gasmodel}
\end{table}%

\subsection{Resolutions}
Total numbers of old stars and gas particles are $3\times 10^6$ ($N_{\rm s}$) and $10^6$ ($N_{\rm SPH}$), and particle masses are $\simeq 1.1\times 10^4 M_\odot$ ($m_{\rm s}$) and $3.2\times 10^3~M_\odot$ ($m_{\rm SPH}$), respectively. With this gas particle mass, gravitational fragmentation of dense gas clumps with $\gtrsim 10^5 M_\odot$ is resolved \citep{Saitoh+2008}.

For calculations of gravitational force, we chose a softening length $\epsilon = 10$ pc. This value is small enough to resolve the three-dimensional structure of a disk galaxy. \citet{Hernquist1987} showed that the softening length $\epsilon$ should be smaller than the mean inter-particle distance $\ell$ at the half mass radius, in order to achieve a good approximation of the gravitational force calculation. For an exponential disk, $\ell \approx 2.63 (N_{\rm s}R_{\rm sd}^{-2}z_{\rm sd}^{-1})^{-1/3}$, we obtain $\ell \approx 35$ pc as an upper limit on the softening length. However, although the smoothing length is allowed to vary time and space, the typical smoothing length is a few times 10 pc in our simulations. Therefore, a spacial resolution in our simulation is estimated as a few 10 pc.

\begin{table}
	\caption{Resolutions}
	\begin{center}
	\begin{tabular}{ll}
	\hline
	Initial number of star particles ($N_{\rm s}$)						& $3 \times 10^6$\\
	Mass of individual star particles ($m_{\rm s}$)					&$1.1 \times 10^4 M_{\odot}$\\ 
	Initial number of SPH particles ($N_{\rm SPH}$)					& $10^6$\\
	Initial mass of individual SPH particles ($m_{\rm SPH}$)	& $3.2 \times 10^3 M_{\odot}$\\
	Gravitational softening length ($\epsilon$)							& $10$ pc\\
	\hline
	\end{tabular}
	\end{center}
	\label{tbl:resolution}
\end{table}%

\section{Results}
\label{sec:results}

\subsection{Evolution of Stellar Bar and Spirals}
\label{sec:spiralevol}

\begin{figure*}[htbp]
	\begin{center}
	\includegraphics[width=.80\textwidth]{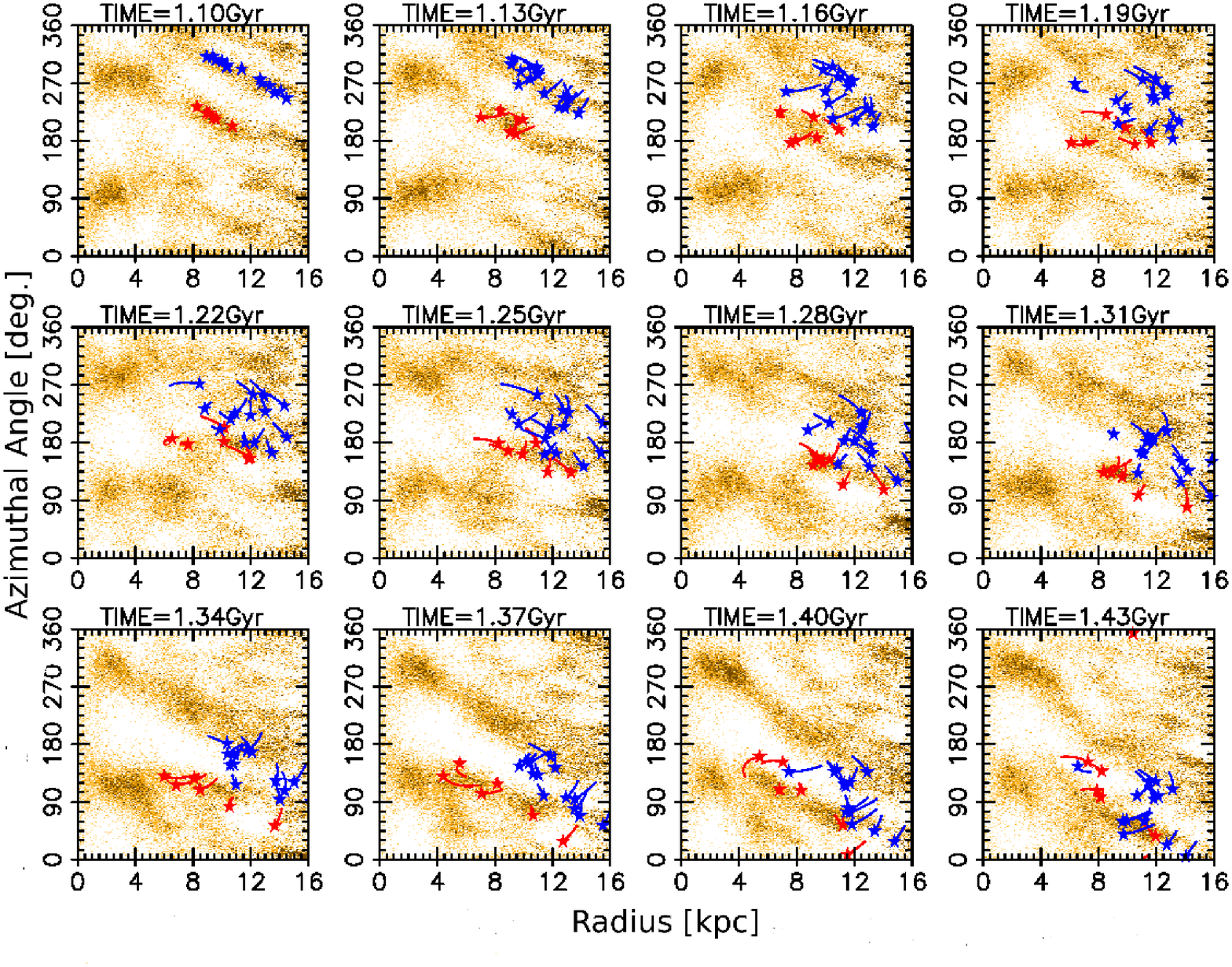}
	\caption{Evolution of normalized stellar density (density contrast) in the plane of azimuthal angle and radial distance, with the angular speed $\Omega_{\rm frame} = 25.6$ km s$^{-1}$ kpc$^{-1}$ which is equal to the pattern speed of the stellar bar. Only the over-dense regions are colored and white regions have the density the below the average value. The average density is obtained by averaging the stellar density for an azimuthal direction at each radius. Red marks and lines represent trajectories of the stars (G1 stars) constituted one spiral arm at $t=1.10$ Gyr, blue ones represent the stars (G2 stars) constituted another spiral arm also at $t=1.10$ Gyr.} 
	\label{fig:StarOrbit}
	\end{center}
\end{figure*}

Figure \ref{fig:StarOrbit} shows the evolution of the stellar disk from $t=1.1$ - $1.5$ Gyr ($t=0$ means the time when the gas component is added to the pre-evolved stellar disk). The bar structure, which is seen as two distinct enhancements at $\phi \sim 90^\circ$ and $270^\circ$, $R < 5$ kpc, stably exists, but spiral arms in the outer disk show very complex behavior. The structures appeared in a co-rotating coordinate of the bar in the plots, we can see that most of the arms drift downward. They drift down faster in the outer region than in the inner region, indicating that they move with the local circular speed. The pattern and number of stellar spiral arms change within a few rotational periods of the galaxy (several 100 Myr). Arm-arm merging and arm break-ups frequently occur.

In order to clarify the evolution of spiral arms in detail, we mark stars in two trailing spiral arms at $t=1.1$ Gyr. Their trajectories are shown in Figure \ref{fig:StarOrbit}. The marked red stars (G1 stars) and the marked blue ones (G2 stars) rotate clockwise on the figure due to epicyclic motions, but their radial oscillation amplitudes are as large as $\sim 2-3$ kpc. From $t=1.1$ to $1.2$ Gyr, the trailing arm marked with the G2 stars spreads as the stars are dispersed. The other arm marked with G1 stars breaks up into two arms: one merges with the outer arm marked with the G2 stars. Then, a new leading arm ($\phi \approx 200^\circ$) forms by $t = 1.19$ Gyr. After $t=1.2$ Gyr, the new leading spiral is wound-up ($t = 1.2 - 1.3$ Gyr) and grows into a single trailing arm at $t=1.4$ Gyr. This new trailing arm consists of the G1 and G2 stars. This is expected in the swing amplification mechanism \citep[][]{Toomre1981, Athanassoula1984}.

As described above, most spiral arms in the model are not steady, but wound-up, and eventually dimmed. A spiral arm often merges with other arms, or breaks up into two or more spirals with coherent motion of the internal stars. Stars in a part of the arms orbit together with the arm in about 1/8-1/4 rotational period at the radius. In other words, the behavior of stellar spirals are closer to `material' arms for the period rather than to quasi-stationary density waves \citep{LinShu1964}, where orbital motion of each star is independent of the kinematics of density waves. Interaction between arms and the splitting of arms are sources of `swing amplification', followed by growth of new trailing spiral arms. As a result, spirals are no longer long-lived, but rather, are recurrent phenomena. Although the theoretical idea of swing amplification is based on a linear approximation \citep[e.g.,][]{GoldreichLynden-Bell1965, JulianToomre1966, Toomre1981, Athanassoula1984}, the mechanism found here is highly non-linear, in a sense that the radial oscillation amplitude of the stars is not small compared to the system size. The recurrent spirals were reported in previous two-dimensional $N$-body simulations \citep{SellwoodCarlberg1984}. However, three-dimensional, high-resolution calculations are necessary to study long-term behavior of the recurrent spirals without being affected by numerical two-body relaxation (J. Baba et al in prep.; M. Fujii et al. in prep.).

\subsection{Kinematics of Young Stars and Star Forming Gas}
\label{sec:pecvel}

\begin{figure*}[htbp]
\begin{center}  
   	\includegraphics[width=.90\textwidth]{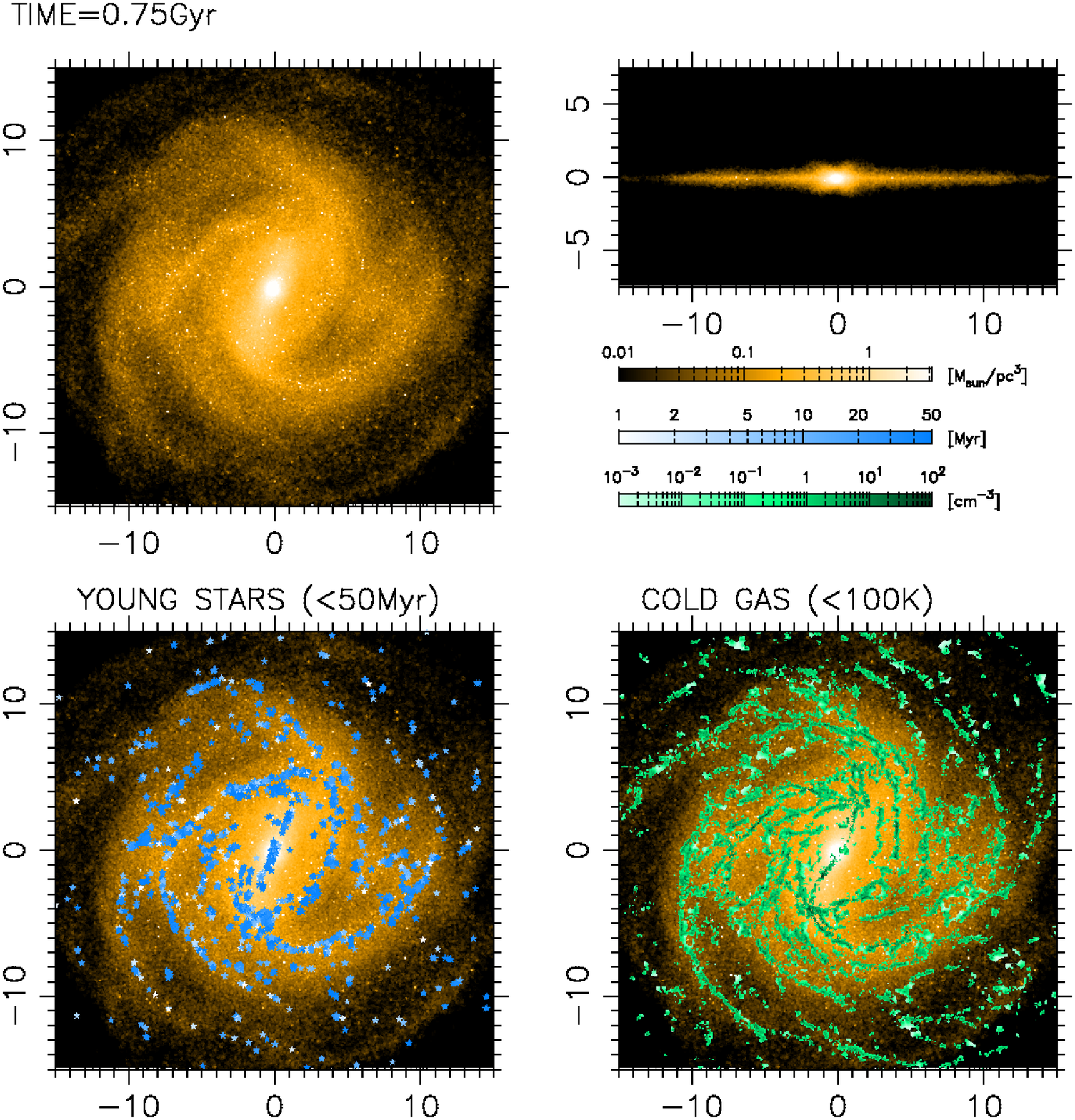}
	\caption{Top left: Distribution of stars projected onto the x-y plane (orange colors indicate stellar density). A major axis of the stellar bar inclines toward $25^\circ$ from a y-axis (vertical axis). Top right: An edge-on view (x-z) of the stellar disk. Bottom left: Young stars, indicated by white-light blue color (age $<50~{\rm Myr}$) overlaid on old stars. Bottom right: Same as the left panel, but cold gas (white-green-black, $T_g < 100$ K) in stead of young stars. The unit length is in kpc.}
	\label{fig:youngstar}
\end{center}
\end{figure*}

Figure \ref{fig:youngstar} shows a snapshot of the system at $t=0.75$ Gyr. A face-on view of the stars is presented in the top left panel. One can see a few grand-design spirals, a bar with the length of about 5 kpc, and a central bulge-like structure. This bulge resembles the bulge of our Galaxy observed by the $COBE$/DIRBE \citep{Dwek+1995} in the edge-on view (top right panel). Clusters of young stars (bottom left panel) roughly trace the background spirals of old stars, and the multi-arm spiral features of the cold gas (bottom right panel) appear. Compared with the background stellar spirals, gas arms are much more filamentary, but they are not hydrodynamical shocks as pointed out by recent hydrodynamic simulations in a fixed spiral potential \citep{Wada2008,Dobbs2008}. Some gaseous spirals are clearly seen even in places where the background stellar spirals are rather weak.  These multi-arm spiral structures of the ISM are morphologically similar to dust emissions in nearby spiral galaxies recently revealed by the {\it Spitzer Space Telescope} \citep{Bendo+2008}.  In previous simulations of disk galaxy formation in which the radiative cooling below $\sim 10^4$ K is not solved \citep[see e.g.,][]{Governato+2008}, the resultant structure of the gas disk is much smoother than that in real galaxies.

\begin{figure*}[htbp]
\begin{center}  
    \includegraphics[width=.90\textwidth]{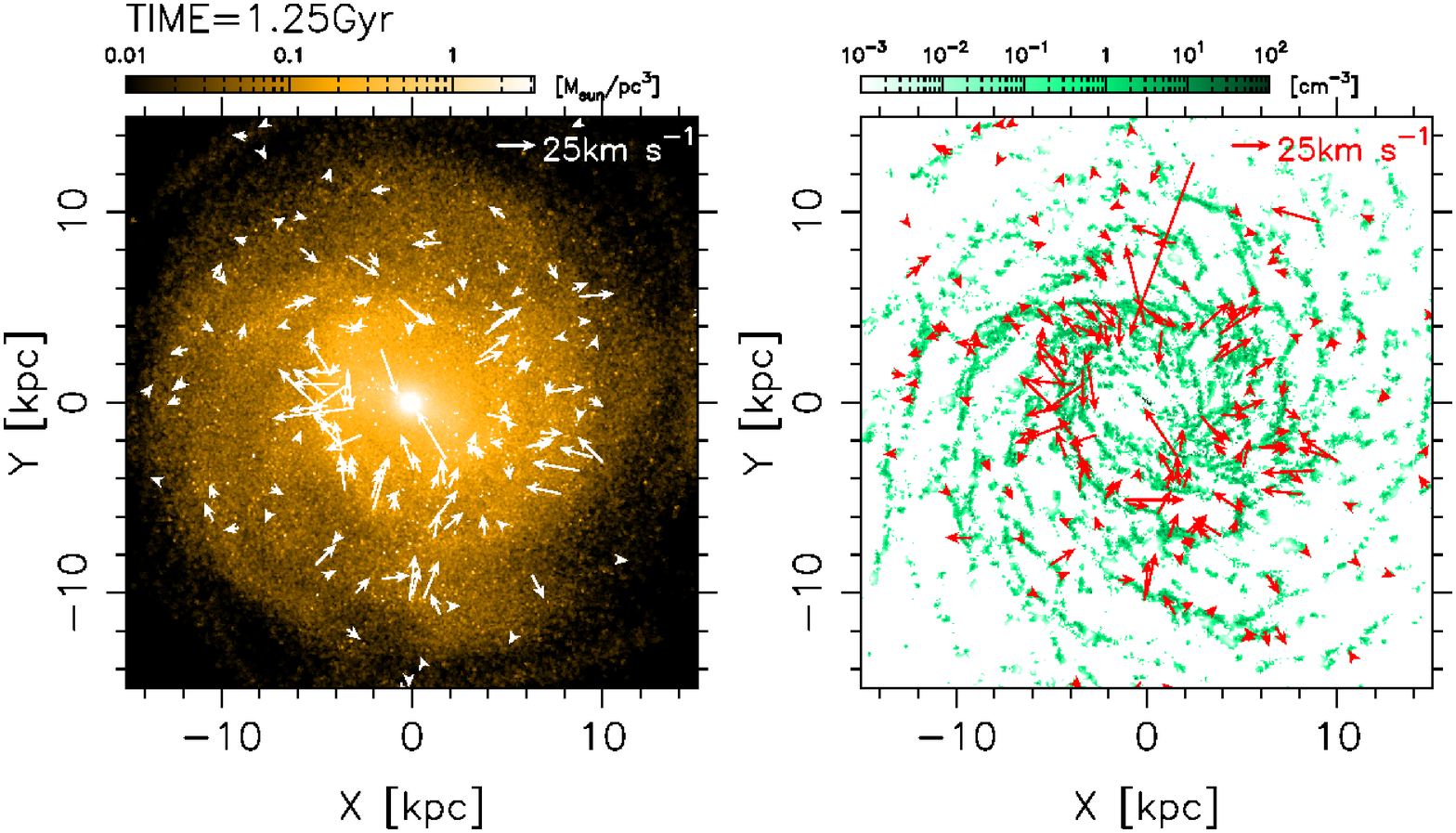}
	\caption{Left: The peculiar velocities of randomly selected young stars (age $< 10~{\rm Myr}$) represented by white arrows overlaid on distribution of stars projected onto the $x-y$ plane (orange colors indicate stellar densities) at $t=1.25~{\rm Gyr}$. The peculiar velocity of a star particle is defined as the deviation from the \emph{true rotation curve}. Right: The peculiar velocities of randomly selected star-forming gas ($n_{\rm H} > 100~{\rm cm^{-3}}$ and $T_{\rm g} < 100$ K) represented by red arrows are overlaid on distribution of cold gas ($T_{\rm g} \leq 100~{\rm K}$, green colors indicate their number densities).  Note that very large peculiar velocities ($\sim 100~{\rm km}~{\rm s}^{-1}$) of a star-forming gas are caused by nearby supernovae explosions. Young stars and star-forming gas within $R=3.0$ kpc are not displayed.}
	\label{fig:PecVelocity}
\end{center}    
\end{figure*}

\begin{figure}[htbp]
\begin{center}  
    \includegraphics[width=.45\textwidth, angle=-90]{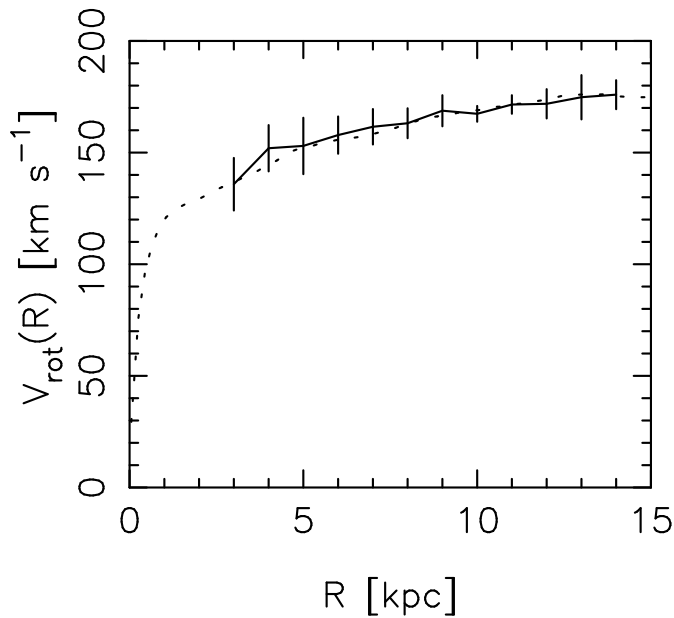}
	\caption{Average rotation velocity of the star-forming gas as a function of a galactocentric radius (solid line). Spacial distribution of these star-forming gas is shown in the right panel of Figure \ref{fig:PecVelocity}. Each bin includes $\sim$ several tens samples, and error bars are the standard deviations of the mean. The dotted line indicates the true rotation curve.}
	\label{fig:RotCurveSFG}
\end{center}    
\end{figure}

The left panel of Figure \ref{fig:PecVelocity} shows peculiar velocities of young stars overlaid on old stars at $t=1.25~{\rm Gyr}$, and in the right panel, peculiar velocities of the high-dense cold gas ($n_{\rm H} > 100$ \ncm3 and $T_{\rm g} < 100$ K, hereafter star-forming gas) are shown on top of the gas density. Here, the peculiar velocity ($v_{\rm pec}$) is defined as deviation from the \emph{true circular rotation} determined by the azimuthally-averaged gravitational field. Many of the star-forming gas clumps and young stars have peculiar velocities of $20 - 30~{\rm km}~{\rm s}^{-1}$, and the directions of the arrows are random. An average rotation speed of the star-forming gas and the true rotation curve are shown in Figure \ref{fig:RotCurveSFG}. The average rotation speed agrees well with the true rotation curve.
Contrary to the recent statement by \citet{Reid+2009b}, in our model, there is no clear tendency that the star-forming gas moves slower than the galactic rotation (see \S \ref{sec:Reid+2009}).

\subsection{Pseudo-observation of Peculiar Velocities of Star Forming Gas}
\label{sec:pecvelsfr}

\begin{figure*}[htbp]
	\begin{center}
	\includegraphics[width=.80\textwidth]{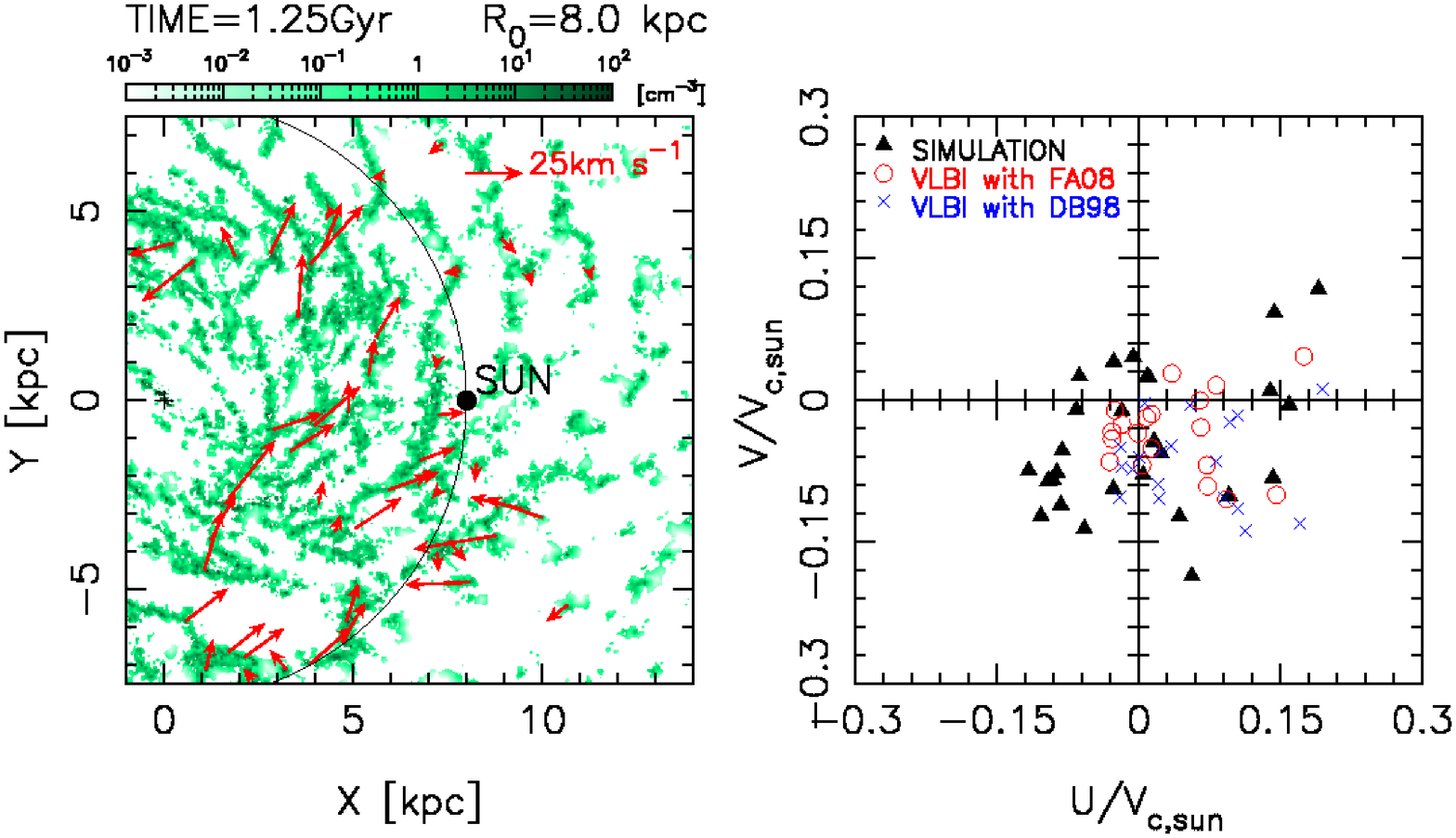}
	\caption{Left: Pseudo-observation of peculiar velocity of randomly selected star-forming gas ($n_{\rm H}>100~{\rm cm^{-3}}$) relative to the ``SUN''. The peculiar velocity is defined by the deviation from the \emph{flat rotation curve} with a galactic rotation velocity at the "SUN". The position angle of the ``SUN'' is assumed to be $25^\circ$ from the major axis of the stellar bar, and its galactocentric radius $R_0 = 8.0$ kpc. The color represents density of cold ($T_{\rm g} < 100~{\rm K}$) gas. The star-forming gas within $R = 3.0$ kpc are not displayed. Right : The Peculiar velocities of the star-forming gas ($n_{\rm H} > 100~{\rm cm^{-3}}$) in $5~{\rm kpc} < R < 10~{\rm kpc}$ on a $U-V$ plane. The U and V are normalized to the galactic rotation velocity at the Sun in the model (triangles) and observations (open circles and crosses). For observational data, the solar motion relative to the LSR is the value of FA09 (open circles), and the value of DB98 (crosses).}
	\label{fig:PecVelocityStat}
	\end{center}
\end{figure*}

The left panel of Figure \ref{fig:PecVelocityStat} shows an example of pseudo-observation of the peculiar velocities of star-forming gas relative to the ``SUN''. The position angle of the ``SUN'' is assumed to be $25^\circ$ from a major axis of the stellar bar, and its galactocentric radius ($R_0$) is assumed to be $8.0$ kpc\footnote{There are many observational suggestions that the solar system is located around $20-30^\circ$ (relative to the Galactic rotation) from the major axis of the bar. The value of $R_0$ is restricted to a range of $7 < R_0 < 9$ kpc, where the observed position-velocity diagram is roughly reproduced.}. A \emph{flat rotation curve} with the circular speed at the ``SUN'' is assumed. The complex distribution of large non-circular motions as seen in the observations (Figure \ref{fig:VLBIobs}) are roughly reproduced. In the simulation, most of the star-forming gas clouds inside the solar circle rotate more slowly than the galactic rotation by $V \sim -10$ to $-20$ \kms (represented by upward arrows), whereas 
they rotate faster (i.e. $V \sim +5$ to $+10$ \kms) in the outer region. These systematic flows are caused by the difference in the true rotation curve from the flat one. Contrary to the simulation, the observational data both inside and outside the solar circle show slow rotations, especially in Figure \ref{fig:VLBIobs}(b).

More quantitative comparison between the observations and the simulation is shown in the right panel of Figure \ref{fig:PecVelocityStat}. Black triangles indicate the star-forming gas shown in left panels. Open circles and crosses indicate the observed star-forming regions, whose peculiar velocities are calculated by assuming the solar motions of FA09 and DB98, respectively (see Table 1). The simulation data points are distributed in $V/V_{\rm c,sun} \sim -0.2$ to $+0.1$ and slightly biased toward $V < 0$. Observed data points, on the other hand, appear in $V/V_{\rm c,sun} \sim -0.15$ to $+0.05$, and apparently they are slower than the galactic rotation (i.e. $V < 0$). This slow rotation appears more clearly in crosses (DB98) than in open circles (FA09), because the solar motion relative to the LSR in crosses (DB98) is $V_{\odot} = 5.25$ \kms, rather than $13.5$ \kms in FA09.

\begin{figure*}[htbp]
	\begin{center}
	\includegraphics[width=.80\textwidth]{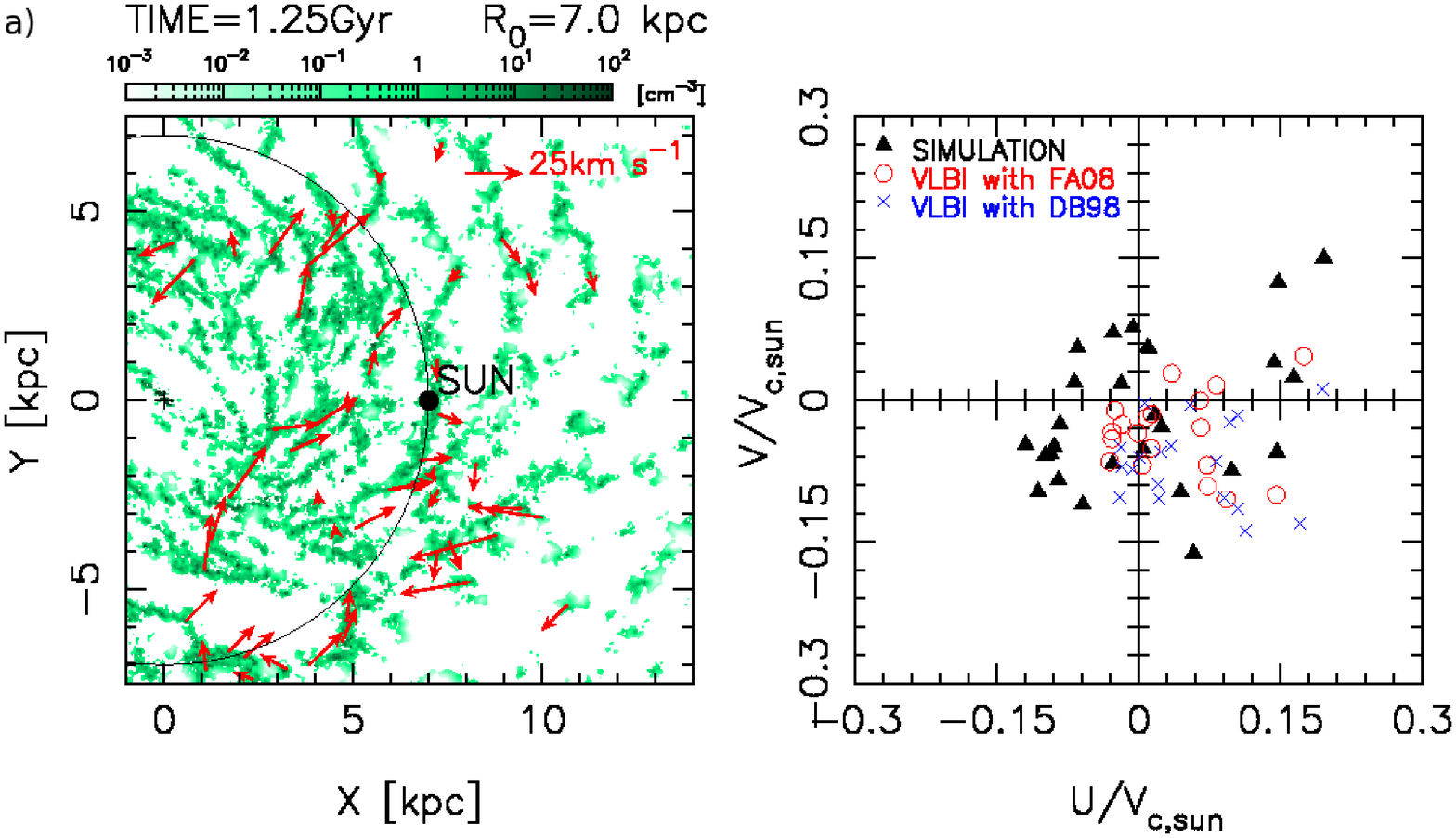}
	\includegraphics[width=.80\textwidth]{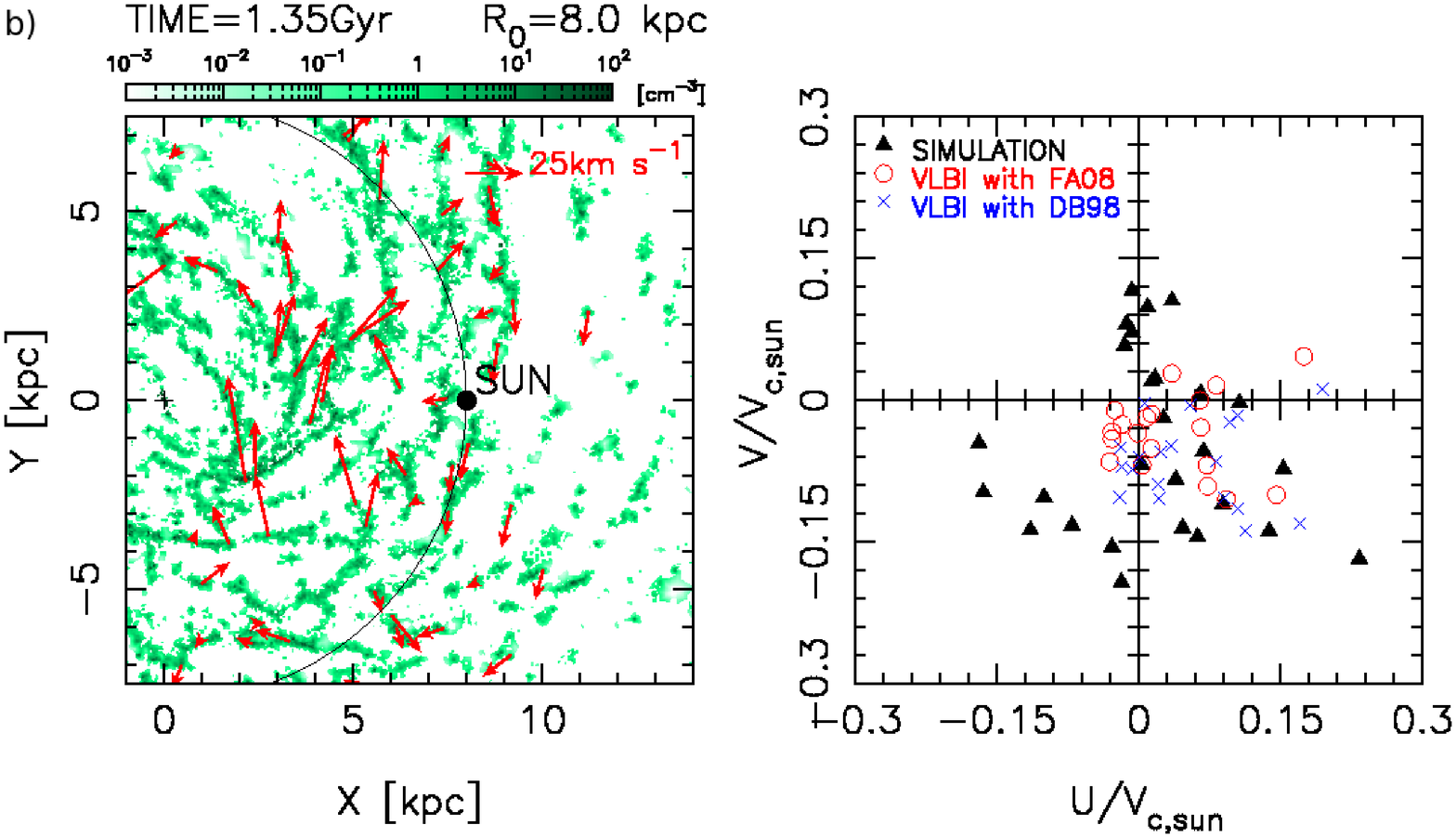}
	\caption{Same as Figure \ref{fig:PecVelocityStat}, but for (a) $R_0=7$ kpc and (b) $t=1.35$ Gyr. }
	\label{fig:PecVelocityStatCmp}
	\end{center}
\end{figure*}

Figure \ref{fig:PecVelocityStatCmp}(a) shows another example of the pseudo-observation from the ``SUN'' at $R_0 = 7$ kpc, it shows no net rotation ($V \approx 0$). The lengths of the arrows inside the solar circle become slightly short, while those outside the solar circle slightly elongated. In contrast, if we place the ``SUN'' further out, $R_0=9$ kpc, the simulation result shows significant bias to $V<0$. Therefore, an average value of $V$ depends on $R_0$. In addition, comparing Figure \ref{fig:PecVelocityStat} with Figure \ref{fig:PecVelocityStatCmp}(b), details of the distribution change through time through the evolution of spiral arms, but the width of the distribution is qualitatively unchanged. 

While the observation shows a clear tendency of inward radial motions ($U > 0$), our simulation results in all cases (Figures \ref{fig:PecVelocityStat}, \ref{fig:PecVelocityStatCmp}(a) and \ref{fig:PecVelocityStatCmp}(b)) do not. One should note that the observed area does not cover the whole galactic disk. In particular, the data points with both $U < 0$ and $V < 0$ on the U-V plane, in the simulation, are the star-forming gas in $X < 5$ kpc and $Y > 0$. This area corresponds to the southern hemisphere for which observation is impossible with either the VLBA or VERA.

\section{Discussion}
\label{sec:discussion}

\subsection{Rotation Speed of Star-Forming Regions}
\label{sec:Reid+2009}

\citet{Reid+2009b} investigated the peculiar motions of high mass star-forming regions in our Galaxy to obtain the Galactic parameters ($R_0$ and $\Theta_0$). They claimed that the star forming regions moves slower than the Galactic rotation. However, this is not the case in our result. As we can see in Figure \ref{fig:RotCurveSFG}, there is no clear tendency that the star-forming gas moves slower than the true rotatin curve.\footnote{\citet{Reid+2009b} adopted the solar motion obtained by DB98 and that they introduced a model in which the star-forming regions move with a certain systematic velocity, $V_{\rm sfr}$, of $-15$ \kms  with respect to the Galactic rotation.} The observed slow rotation \citep{Reid+2009b} is probably caused by an uncertainty of the solar motion relative to the LSR, especially, $V_{\odot}$. If we compare with the right and left panels in Figure 1, it is clear that the solar motion of FA09 reduces the amount of slow rotation. If \citet{Reid+2009b} adopt the solar motion of FA09 instead of DB98, $V_{\rm sfr}$ could be smaller in their analysis. We have to note that ``SUN'' for an observer might have a large uncertainty in the peculiar motion. Therefore the peculiar motions of the star-forming regions could have a bias with respect to the Galactic rotation curve. Recently, \citet{McMillanBinney2009} suggests that the value of $V_{\odot}$ determined in these papers may be underestimated by $\sim 6-7~{\rm km}~{\rm s}^{-1}$. Their value is close to the value of FA09 adopted in our paper. Our result is consistent with the argument of\citet{McMillanBinney2009}.

\subsection{Origin of Large Peculiar Motions}
\label{sec:origin}
In order to discuss the origin of the large peculiar motions of the star-forming gas, we compare three runs (Runs FID, ISO, and NSF): In Run FID, which is presented in previous sections, the gas can cool down to $20$ K, and star formation from the dense and cold gas ($n_{\rm H} > n_{\rm th} = 100$ cm$^{-3}$ and $T_{\rm g} < T_{\rm th} = 100$ K) is included. In Run ISO, the gas can cool down to $10^4$ K, and star formation from the gas with $n_{\rm H} > 0.1$ cm$^{-3}$ and $T < 15000$ K is included. In Run NSF, the gas can cool down to $20$ K as in Run FID, but this run included neither star formation nor SNe.

\begin{figure}[htbp]
	\begin{center}
	\includegraphics[width=.45\textwidth, angle=-90]{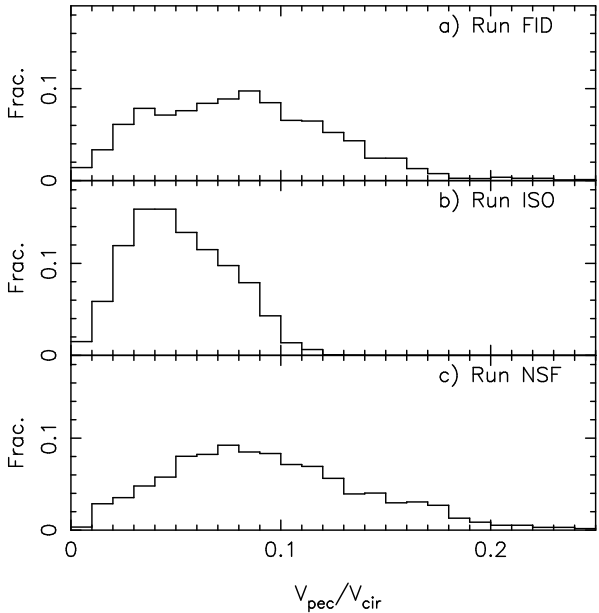}
	\caption{Distribution of the ratio of peculiar velocity to circular velocity, $v_{\rm pec}/V_{\rm cir}$, for star-forming gas in 5 kpc $<$ R $<$ 10 kpc at $t= 1.30$ Gyr. The peculiar velocity ($v_{\rm pec}$) is defined as the deviation from the local circular velocity. a) Run FID, b) Run ISO, and c) Run NSF. See text in \S\ref{sec:origin} for the detailed explanation about these Runs.}
	\label{fig:PecVelStat}
	\end{center}
\end{figure}

\begin{figure}[htbp]
	\begin{center}
	\includegraphics[width=.45\textwidth, angle=-90]{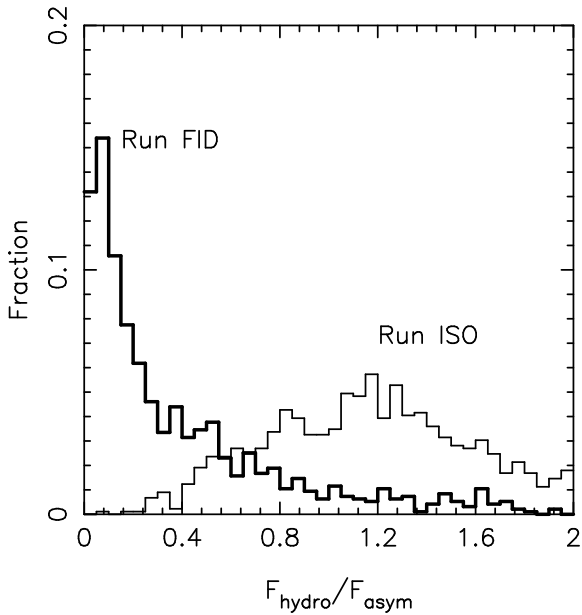}
	\caption{Distribution of the ratio of hydrodynamical force ($F_{\rm hydro}$) to asymmetric gravitational force ($F_{\rm asym}$) for star-forming gas shown in Figure \ref{fig:PecVelStat}. Thick and thin curves indicate the results of Runs FID and ISO, respectively.}
	\label{fig:PecForce}
	\end{center}
\end{figure}

Figure \ref {fig:PecVelStat}b shows the histogram of $v_{\rm pec}/V_{\rm cir}$ of the star-forming gas in Run ISO. They have peculiar velocities of $\sim 5\%$ of the circular velocities, which cannot explain the observed large peculiar velocities of $\sim 10 \%$ of the circular velocity. Figure \ref {fig:PecVelStat}a shows the result of Run FID, where the star-forming gas has $\sim 2$ times larger peculiar velocities than those in Run ISO. In order to understand what causes this difference, the distribution of a ratio of the pressure gradient ($F_{\rm hydro}$) to the non-axisymmetric component of the gravitational field ($F_{\rm asym}$) was investigated as shown in Figure \ref{fig:PecForce}. In Run ISO, many star-forming gases are distributed around $F_{\rm hydro} \sim F_{\rm asym}$. However, most of the star-forming gases in Run FID are distributed $F_{\rm hydro} < 0.1 \times F_{\rm asym}$. Therefore, the non-circular motion of star-forming gas in Run ISO is not damped by hydrodynamic pressure in the case of Run FID, and so, they can interact with transient stellar spirals via mechanism such as swing amplification (see \S\ref{sec:spiralevol}).

Finally, we compare Run FID with Run NSF. The histogram of $v_{\rm pec}/V_{\rm cir}$ of the star-forming gas in Run NSF is shown in Figure \ref {fig:PecVelStat}c.  The distribution is almost the same as that in Run FID, suggesting that the acceleration by the SNe is not the main contributor in causing the large peculiar motions of the star-forming gas.

\subsection{Kinematic Distances and Face-on View of our Galaxy}
\label{sec:kderror}

Since the pioneering work by \citet{Oort1958} there have been many attempts to construct the large-scale spiral structure of our Galaxy using the conventional ``kinematic distance'' method as a first approximation. The kinematic distance method is based on assumptions that the orbits are not very different from circular with an angular orbital speed decreasing monotonically as a function of galactocentric radius. Nearby major spiral arms were identified by applying this method to various sources such as HI, CO, and HII regions \citep[e.g,][]{Oort1958, NakanishiSofue2003, NakanishiSofue2006, GeorgelinGeorgelin1976}.

\begin{figure*}[htbp]
\begin{center}  
    \includegraphics[width=.50\textwidth, angle=-90]{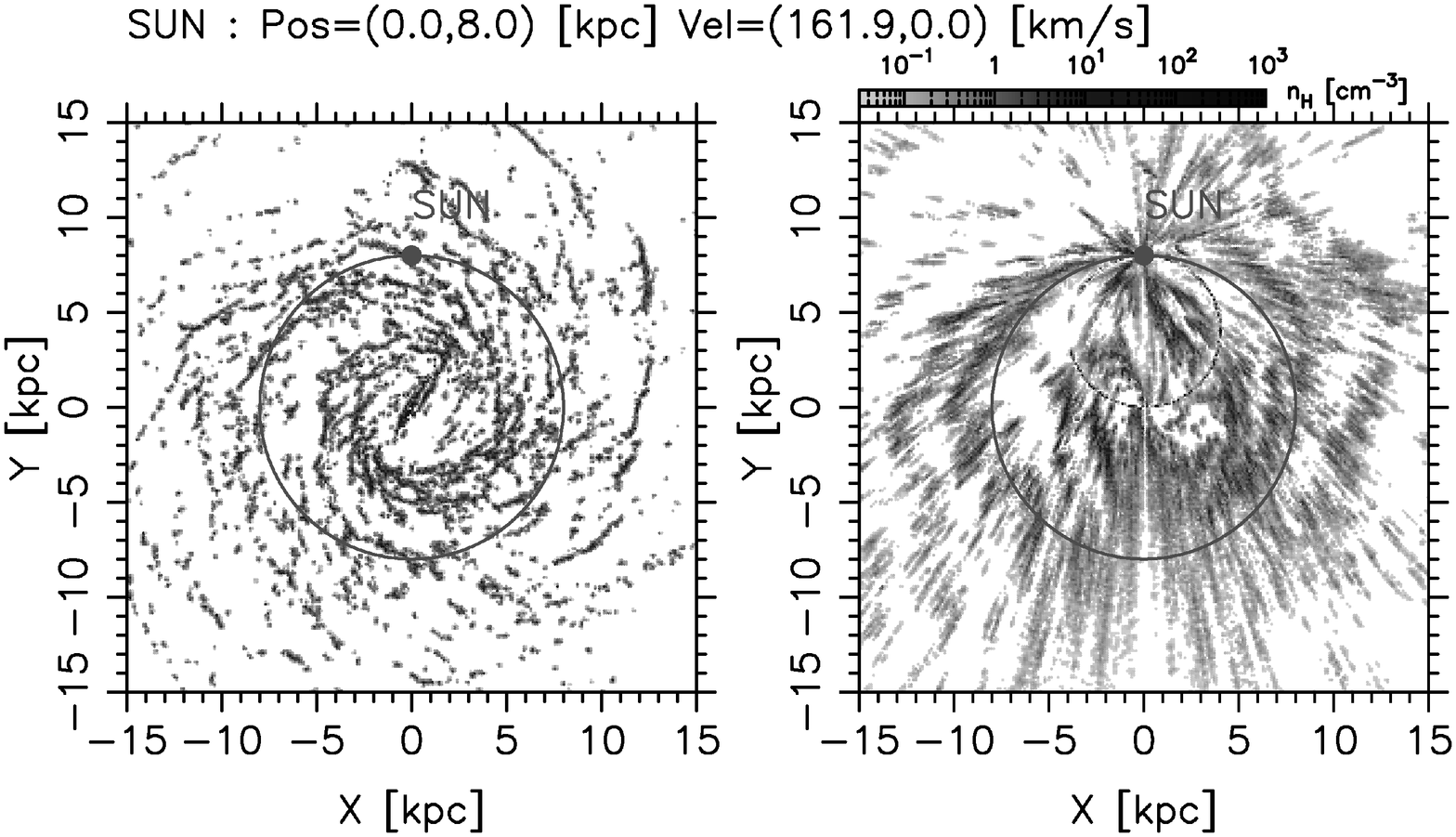}
	\caption{Left: Distribution of the cold gas ($T_{\rm g} \sim 100$ K) in the numerical model at $t=2.0~{\rm Gyr}$. Right: Same as the left panel, but reconstructed using a conventional method, i.e. the kinematic distance determined by the line-of-sight velocity ``observed'' from the position of ``SUN'' (marked a red filled circle), assuming that the all gases rotate circularly.  The position angle of the ``SUN'' is assumed to be $25^\circ$ from the major axis of the stellar bar (see Figure  \ref{fig:youngstar}) and its galactocentric radius is assumed to be $8.0$ kpc. }
	\label{fig:HiMap}
\end{center}    
\end{figure*}

\begin{figure}[htbp]
\begin{center}
	\includegraphics[width=.45\textwidth, angle=-90]{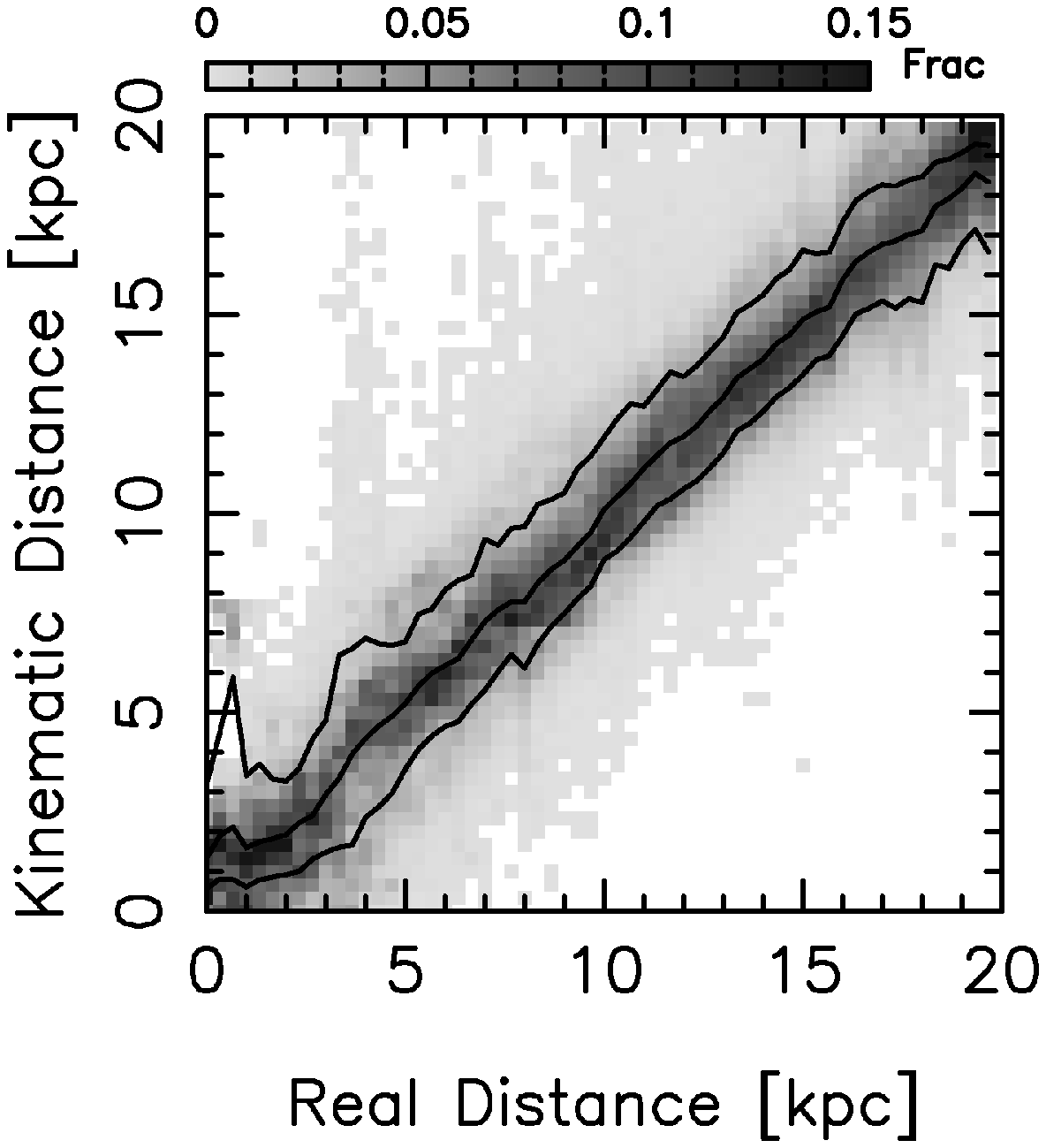}
	\caption{Kinematic distance obtained assumed circular rotation vs. real distance from the observer (see Figure \ref{fig:HiMap}). In the worst cases, kinematic distances differs from the real ones as large as $\sim 4-6$ kpc for any distances. Three solid lines indicates the average kinematic distance, and 1-$\sigma$ errors in each real distance bin.}
	\label{fig:KinmDistanceError}
\end{center}
\end{figure}

The typical peculiar velocity of $20-30~{\rm km}~{\rm s}^{-1}$, which is suggested by our numerical model and recent VLBI observations, might have a strong impact on how we see our Galaxy. The two panels of Figure \ref{fig:HiMap} demonstrate the difference between the true gas distribution in our model galaxy and the view obtained using the kinematic distance technique, where the velocity field of gas is used to derive the distance from the ``SUN'' \footnote{\citet{Combes1991} showed a similar comparison based-on their `cloud-fluid' model, but a fixed stellar bar potential was assumed.}. Since our aim here is to demonstrate how the large non-circular motions affect the estimations in kinematic distances, the errors caused by the distance ambiguity in the inner galaxy have to be in a minimum. Therefore, we adopted the same method as \citet{Gomez2006}: based on the real distribution of the ISM, we first figure out whether a target gas element (particle) is located at the near or far side, and then we derive distance of the element using its line of sight velocity assuming the circular rotation. The reconstructed map roughly reproduce the real distribution of the gas, but it shows many clear `spur'-like structures towards the Sun, similar to those seen in the previously published HI and CO maps of our Galaxy \citep{Oort1958,NakanishiSofue2003,NakanishiSofue2006}. Most of these structures are not real, but they are the result of the large non-circular motions of the gas clouds as well as the non-axisymmetric flow caused by the bar. In addition, outside the solar circle the arms in the left half of the figure have a pitch angle larger than that of the real arm, while in the right half the pitch angles in the kinematic map are smaller than the real ones. This asymmetry is also visible in the maps constructed from the observations, suggesting that the motions of gas clouds in our Galaxy actually have large non-circular motions, similar to what was obtained in the present simulations, and that the location of the Sun relative to the major axis of central bar has an offset. More quantitatively, Figure \ref{fig:KinmDistanceError} shows that the kinematic distances are well correlated with the real ones, but there are large errors with the typical value of $\sim \pm (2-3)$ kpc in the kinematic ones from the real ones. 

If we apply the kinematic distance method to the objects associated with the spiral arms, we must consider large non-circular motions. There is an approach which involves the modeling of non-circular motions of the gas instead of assuming a circular motion \citep[e.g.][]{FosterMacWilliams2006, Gomez2006, Pohl+2008}. Recently, \citet{Pohl+2008} recovered a ``real'' 3-D distribution of molecular gas from CO observations using a hydrodynamic simulation. They used a probabilistic method to match the observational CO gas distribution to the model prediction along the line-of-sight, instead of assuming a circular motion for the underlying kinematic model. However, the velocity field of the ISM in previous works \citep[][]{Gomez2006, Pohl+2008} was probably much smoother than the real one because the ISM was treated as an isothermal gas ($T_{\rm g} \sim 10^4$ K) and the stellar bar was treated as a fixed potential. As discussed in \S \ref{sec:origin}, the cold gas clumps can have large non-circular motions through the non-linear gravitational interaction between them and transient stellar spirals. In order to recover the gas distribution in our Galaxy, it would be necessary to use a more complex velocity field as naturally expected for the multi-phase ISM in a live stellar disk.
 
\section{Summary and Conclusion}
\label{sec:sum}
High-resolution, $N$-body+hydrodynamical simulation in which the multi-phase ISM and star formation are self-consistently taken into account revealed the kinematics of star-forming regions in a barred-spiral, like our Galaxy. We found that stellar and gas arms are not steady but are transient and recurrent. This transient nature of the spiral arms may contribute to the large and complex non-circular motions of star-forming regions, as found in recent VLBI observations. 

We found that the peculiar motions of the Galactic star forming regions are qualitatively consistent with the motions of the star forming gas in our model. Even though the kinematic distance technique roughly follows the real distance, each spiral arm is stretched as large as $4 - 6$ kpc along the line of sigt in the reproduced map. This is because of the assumption in the kinematic distance, i.e. the pure circular rotation.

The results shown here suggest that systematic astrometric measurements of objects in the galactic disk are essential in order to understand the structure of our Galaxy. The VLBI astrometry and future space missions for infrared astrometry such as the original plan of Japanese JASMINE \citep{Gouda+2006} or even the astrometry of nearby (a few kpcs) disk stars which should be possible with the GAIA \citep{Perryman+2001} will give us a completely new picture of the spiral arms of our Galaxy, with the help of numerical simulations like the ones presented here, or ultimately cosmological galaxy formation simulations with sufficiently high enough resolutions.

\acknowledgments
The authors are grateful to the anonymous referee for his/her constructive comments. We also thank F. Combes, and D. Lynden-Bell for their valuable comments. Calculations and visualization were performed by GRAPE-7 and  Cray XT-4 in CfCA, National Astronomical Observatory of Japan. This study is supported by the Molecular-Based New Computational Science Program, NINS. TRS is financially supported by a Research Fellowship from the Japan Society for the Promotion of Science for Young Scientists.

\bibliographystyle{apj}
\bibliography{ms}

\end{document}